Sample Size Calculations in Simple Linear Regression: Trials and Tribulations

Tianyuan Guan, M. Khorshed Alam, M. Bhaskara Rao

*Department of Environmental Health, Division of Biostatistics and Bioinformatics, University of Cincinnati, 160 Panzeca Way, Cincinnati OH 45267 – 0056*

Abstract
The problem tackled in this paper is the determination of sample size for a given level and power in the context of a simple linear regression model. At a technical level, the simple linear regression model is a five-parameter model. It is natural to base sample size calculations on the least squares' estimator of the slope parameter of the model. Nuisance parameters such as the variance of the predictor X and conditional variance of the response Y create problems in the calculations. The current approaches in the literature are not illuminating. One approach is based on the conditional distribution of the estimator of the slope parameter given the data on the predictor X. Another approach is based on the sample correlation coefficient. We overcome the problems by determining the exact unconditional distribution of the test statistic built on the estimator of the slope parameter. The exact unconditional distribution alleviates difficulties to some extent in the computation of sample sizes. On the other hand, the test based on the sample correlation coefficient of X and Y avoids the problems besetting the test based on the slope parameter. However, we lose intuitive interpretation that comes with the slope parameter. Surprisingly, we see that the sample size that comes from the correlation test works in synchronization with the one that comes from the test built upon the slope parameter in a broad array of settings.

1. INTRODUCTION

Multiple regression is one of the core methodologies in statistics.   Power computation and sample size determination have become integral part of many research proposals submitted for funding.  Horton and Switzer (2005) report that 51% of research articles published in the New England Journal of Medicine during 2004-05 have Multiple Regression as one of the methods used.  The figure for power analysis was 39%.  In this paper, we focus on power computation in the context of simple linear regression. Is there a problem in this setting? Regression methodology has a long history dating back to Galton (1885). Funding agencies such as NIH (National Institute of Health) have been demanding sample size calculations in the proposals since its inception. We point out difficulties in this setting. See Ryan (2013).

To provide a background and rationale of our work, we first review how sample size is determined in the literature under the simple linear regression paradigm. A simple linear regression model has two entities X, the predictor, and Y, the response variable. The model is stated as

$$Y|X \sim N(\beta_0 + \beta_1 X, \sigma^2)$$

for some $\beta_0, \beta_1$ and $\sigma^2 > 0$. The hypothesis of interest is $H_0: \beta_1 = 0$ against the alternative $H_1: \beta_1 \neq 0$. What should be the required sample size, $n$, for a given level of significance $\alpha$, power $1 - \beta$, and at the alternative value A of $\beta_1$. Let $(X_1,Y_1), (X_2,Y_2), ..., (X_n,Y_n)$, n independent realizations of (X,Y), be a potential sample for the testing problem. Let $\hat{\beta}_1$ be the least squares estimator of β1, i.e.,

$$\hat{\beta}_1 = \frac{S_{XY}}{S_{XX}}$$

where

$$S_{XY} = \sum_{i=1}^{n}(X_i - \bar{X})(Y_i - \bar{Y})$$

and

$$S_{XX} = \sum_{i=1}^{n}(X_i - \bar{X})^2.$$

Let RSS be the residual sum of squares, i.e.,

$$RSS = \sum_{i=1}^{n}\left[(Y_i - \bar{Y}) - \hat{\beta}_1(X_i - \bar{X})\right]^2$$

For testing the null hypothesis H₀, we use the test statistic:

$$T = (\hat{\beta}_1\sqrt{S_{XX}})/(\sqrt{RSS/(n-2)}) .$$

Under the null hypothesis, conditioned on the X-data, $T$ has a $t$-distribution with $n - 2$ degrees of freedom and under the alternative hypothesis $\beta_1$= A, $T$ has a non-central $t$-distribution with degrees of freedom (n – 2), and non-centrality parameter.

$\lambda = \frac{A^2 S_{XX}}{\sigma^2}$.

We reject null hypothesis if and only if $|T| > t_{1-\frac{\alpha}{2},n-2}$ where $t_{1-\frac{\alpha}{2},n-2}$ is such that the area to the left of $t_{n-2}$ – student's t – curve is $1 - \frac{\alpha}{2}$. The power formula is given by

$$\text{Power}(A) = Pr(\text{Reject } H_0|\beta_1 = A) = Pr\left(|T| > t_{1-\frac{\alpha}{2},n-2}|\beta_1 = A\right)$$

We can set power = 1-β and solve for n. This would work as long as we know what $\lambda = \frac{A^2 S_{XX}}{\sigma^2}$ is. This would require knowledge of the alternative value of β1, $\sigma^2$, and Sxx.

Critically, we should know for each n what Xi values are, so that $(1/n)\sum_{i=1}^{n}(X_i - \bar{X})^2$ is a constant. Equivalently, one should spell out what λ is. This is a tall order. Adcock (1997) recognized these problems. Some software and textbooks do assume that $(1/n)\sum_{i=1}^{n}(X_i - \bar{X})^2$ is known and a constant. For example, the software PASS and nQuery do proceed this way. More realistically sampling is done on (X, Y) simultaneously. In that case, we cannot use T. To accommodate this scenario, we proceed with deriving the exact unconditional distribution of a variant of T. We will assume that the predictor X has a normal distribution with mean $\mu_X$ and standard deviation $\sigma_X$.

The five-parameter model now is:
$$Y|X \sim N(\beta_0 + \beta_1 X, \sigma^2)$$

$$X \sim N(\mu_X, \sigma_X^2).$$

We will use a variant of T by utilizing the standard unbiased estimator of $\sigma_X^2$. We use the same notation.
$$T = \frac{\hat{\beta}_1 * \hat{\sigma}_X}{\hat{\sigma}} \quad (1)$$

where $\hat{\sigma}^2 = \frac{RSS}{n-2}$ and $\hat{\sigma}_X^2 = \frac{S_{XX}}{n-1}$.

As an alternative to (1), one can use the correlation coefficient ρ between Y and X. The correlation has an explicit formula.
$$\rho = \frac{\beta_1 * \sigma_X}{\sqrt{\beta_1^2 \sigma_X^2 + \sigma^2}}$$

The statement $\beta_1 = 0$ is equivalent to $\rho = 0$. One can use the sample correlation coefficient $\hat{\rho}$ to build a test for testing $\rho = 0$. The null distribution of the test statistic based on $\hat{\rho}$ is elegant and its non-null distribution is tractable. We can carry out power calculations of the test based on $\hat{\rho}$ with a great degree of ease. However, many researchers prefer to use the test based on $\hat{\beta}_1$, because $\beta_1$ has an easy and physical interpretation (Ryan, 2013). Technically, we cannot use the sample size calculated based on $\hat{\rho}$ with the test based on $\hat{\beta}_1$. Really? We show that there is a good agreement in sample sizes calculated in the environment of both the tests in a broad array of settings. Test-hopping is feasible, after all!

The paper is organized as follows. In Section 2, we provide a literature review and outline main results. In Section 3, we derive the unconditional distribution of T under the null hypothesis. In Section 4, we calculate critical values using the main result of Section 3. In Section 5, we lay out the sample sizes required for a given level, power, and effect size. In Section 6, we take up the problem of finding sample sizes based on the correlation. In Section 7, we contrast the sample sizes. In Section 8, the paper is summarized, and conclusions are drawn.

2. Literature Review and main results outlined

Ryan has pointed out difficulties in power calculations in the environment of simple linear regression (Ryan, 2013). The problem is how we handle the predictor X. Adcock has looked at some possible scenarios (Adcock, 1997). One scenario is that the investigator knows the $X_i$-values (deterministic) for every sample size n. In such a case, the test statistic

$$(\hat{\beta}_1 \sqrt{S_{XX}})/(\sqrt{RSS/(n-2)}) \qquad (2)$$

is eminently usable for power calculations. Its (conditional) null and non-null distributions have been worked out explicitly. The software PASS requires specification of α (size), alternative value of $\beta_1$, $S_{XX}$, sample size, and σ (error standard deviation) for power calculation. The software nQuery also follows the same route (Dupont and Plummer, 1998; Draper and Smith, 1985; Hsieh, Bloch, and Larsen, 1998; Maxwell, 2000; Thigpen, 1997; Wetz, 1984).

A more natural scenario is that we sample (X,Y) simultaneously. We have to handle the test statistic (2) carefully. We need its unconditional distribution, which seems to be intractable at the outset. An alternative to the test statistic (2), we can build a test based on the sample correlation coefficient, $\hat{\rho}$. The null and non-null distributions of the test have been worked out explicitly. In our consulting work, many researchers prefer to use the test based on $\hat{\beta}_1$. It is a choice between causality and association (Ryan, 2013; Cohen,1988; SAS; Gatsonis and Sampson, 1989; Krishnamoorthy and Xia, 2008).

In our research here, we will stick to $\hat{\beta}_1$. We use the test statistic T in (1).

We determine the distribution of T under the null hypothesis $\beta_1$ = 0. As a matter of fact, we show that

$$T^2 \sim \frac{(n-2)}{(n-1)} * \frac{W_1 W_4}{W_2 W_3},$$

where $W_1 \sim \chi_1^2$, $W_2 \sim \chi_{n-1}^2$, $W_3 \sim \chi_{n-2}^2$ and $W_4 \sim \chi_{n-1}^2$, with $W_i$ s being mutually independent. We use this result to obtain the critical values of the test based on T, for given levels. For power and sample size computations, we need the distribution of T for a given value of $\beta_1$. The distribution also depends on $\sigma_X^2$ and $\sigma^2$. It turns out that the distribution depends only on turns out that the distribution depends only on λ = $\beta_1 \frac{\sigma_X}{\sigma_\varepsilon}$, which we can deem as the effect size. The specification of λ facilitates computation of power. In spite of all these deliberations, no magic formula for power surfaces. Knowing the distribution of $T^2$ under λ eases the pain a little bit.

One might skip all this hard work by rooting for the sample correlation coefficient ρ. One could use the statistic

$$T_1 = \frac{\sqrt{(n-2)} * \hat{\rho}}{\sqrt{1-\hat{\rho}^2}} \qquad (3)$$

for testing and power computations. The null and non-null distributions of $T_1$ have been well-documented. One needs to spell out the alternative value of ρ for power computation.

Can one use the sample size that emanates from here with the test based on $\hat{\beta}_1$ after matching the alternative value of ρ with the effect size λ? Is test-hopping legitimate? Surprisingly, it works. This will be discussed in Section 7.

### 3. **Distributional Results**

In this section, we will derive the probability density function of T of (1) unconditionally. The following series of steps will give the desired result.

1. Given $X_1, X_2, \ldots, X_n$, $\hat{\beta}_1$ has a normal distribution with mean $\beta_1$ and variance $\frac{\sigma^2}{S_{XX}}$ and $\hat{\beta}_1$ and $RSS$ are independent.
2. Unconditionally, $\frac{RSS}{(n-2)\sigma^2} \sim \chi^2_{n-2}$.
3. $\frac{S_{XX}}{\sigma_X^2} \sim \chi^2_{n-1}$.
4. $RSS$ and $S_{XX}$ are independent.

The random variable we entertain is $T = \frac{(\hat{\beta}_1 - \beta_1)\hat{\sigma}_X}{\hat{\sigma}_\varepsilon}$ for a given value of $\beta_1$.
The goal now is to get the unconditional distribution of $T$.

The joint distribution of $\hat{\beta}_1$ and $S_{XX}$:

$$g(\hat{\beta}_1, S_{XX}) = \frac{\sqrt{S_{XX}}}{\sqrt{2\pi}\sigma} * e^{-\frac{S_{XX}}{2\sigma^2}(\hat{\beta}_1 - \beta_1)^2} * \frac{1}{\Gamma\left(\frac{n-1}{2}\right) * 2^{\frac{n-1}{2}}} e^{-\frac{S_{XX}}{2*\sigma_X^2}} \left(\frac{S_{XX}}{\sigma_X^2}\right)^{\frac{n-1}{2}-1} \left(\frac{1}{\sigma_X^2}\right)$$

$-\infty < \hat{\beta}_1 < \infty, \; 0 < S_{XX} < \infty$

The (unconditional) marginal density of $\hat{\beta}_1$ is given by

$$f(\hat{\beta}_1) = \frac{1}{2^{\left(\frac{1}{2}\right)}2^{(n-1)/2}\sqrt{\pi}\Gamma\left(\frac{n-1}{2}\right)\sigma*(\sigma_X^2)^{\frac{n-1}{2}}} \int_0^\infty S_{XX}^{\frac{n}{2}-1} e^{-\frac{S_{XX}}{2}\left(\frac{1}{\sigma_X^2} + \frac{(\hat{\beta}_1 - \beta_1)^2}{\sigma^2}\right)} dS_{XX}$$

$$= \frac{\Gamma\left(\frac{n}{2}\right)}{\sqrt{\pi}\Gamma\left(\frac{n-1}{2}\right)\sigma * (\sigma_X^2)^{\frac{n-1}{2}}} \left(\frac{1}{\frac{1}{\sigma_X^2} + \frac{(\hat{\beta}_1 - \beta_1)^2}{\sigma^2}}\right)^{\frac{n}{2}}$$

$$= \frac{\sigma_X}{B\left(\frac{1}{2}, \frac{n-1}{2}\right)\sigma} * \frac{1}{\left(1 + \frac{(\hat{\beta}_1 - \beta_1)^2 \sigma_X^2}{\sigma^2}\right)^{\frac{n}{2}}}, \quad -\infty < \hat{\beta}_1 < \infty$$

Some properties of this density are clear to observe. For example, the distribution is symmetric around the true value $\beta_1$. If n=2, the distribution is Cauchy. In addition, $\frac{\sigma_X}{\sigma}(\hat{\beta}_1 - \beta_1)\sqrt{n-1} \sim t_{n-1}$. Further, if n > 3, unconditionally,

1. $E(\hat{\beta}_1) = \beta_1$ and $\text{Var}(\hat{\beta}_1) = \frac{1}{n-3} * \frac{\sigma^2}{\sigma_X^2}$

2. In the conditional set-up,
$$E(\hat{\beta}_1|X_1, X_2, \ldots, X_n) = \beta_1$$
$$\text{Var}(\hat{\beta}_1|X_1, X_2, \ldots, X_n) = \frac{\sigma^2}{S_{XX}}$$

3. The random variable $U = \frac{(\hat{\beta}_1 - \beta_1)\sigma_X}{\sigma}$ has the probability density function:
$$f(U) = \frac{1}{B\left(\frac{1}{2}, \frac{n-1}{2}\right)} * \frac{1}{(1+U^2)^{\frac{n}{2}}}, -\infty < U < \infty.$$

4. It follows that $U^2 \xrightarrow{d} \frac{W_1}{W_2}$, where $W_1 \sim \chi_1^2$ and $W_2 \sim \chi_{n-1}^2$, with $W_1$ and $W_2$ being independent.

5. Exact distribution of $T$: note that $\frac{(\hat{\beta}_1 - \beta_1)}{\sigma_\varepsilon}$ and $\hat{\sigma}_X^2$ are independent

6. $T^2 = \frac{(\hat{\beta}_1 - \beta_1)^2 \hat{\sigma}_X^2}{\hat{\sigma}_\varepsilon^2} = \frac{(\hat{\beta}_1 - \beta_1)^2 \sigma_X^2}{\sigma_\varepsilon^2} * \frac{\sigma_\varepsilon^2}{\hat{\sigma}_\varepsilon^2} * \frac{\hat{\sigma}_X^2}{\sigma_X^2} \xrightarrow{d} \frac{W_1}{W_2} * \frac{n-2}{W_3} * \frac{W_4}{n-1}$, where $W_3 \sim \chi_{n-2}^2$ and $W_4 \sim \chi_{n-1}^2$, and with $W_1, W_2, W_3,$ and $W_4$ being independent.

In short, $T^2 \sim \frac{(n-2)}{(n-1)} * \frac{W_1 W_4}{W_2 W_3}$.

7. It follows that:
$$E(T^2) = \frac{(n-2)}{(n-3)(n-4)}$$

**4. Critical Values**

The null hypothesis is: $H_0 : \beta_1 = 0$ and the alternative is $H_1 : \beta_1 \neq 0$.
Test Statistic:
$$T = \frac{\hat{\beta}_1 \hat{\sigma}_X}{\hat{\sigma}}$$

Test: Reject the null hypothesis if and only if $|T| > C$, where the critical value C depends on the sample size n and level of significance α. We denote the critical value by $C_{n,\alpha}$.
The null distribution of $T^2$ has been determined in Section 3.
The critical value $C_{n,\alpha}$ satisfies the equation:
$$\alpha = Pr\left(\left|\frac{\hat{\beta}_1 \hat{\sigma}_X}{\hat{\sigma}}\right| > C_{n,\alpha} | H_0 : \beta_1 = 0\right) = Pr\left(\frac{\hat{\beta}_1^2 \hat{\sigma}_X^2}{\hat{\sigma}^2} > C_{n,\alpha}^2 | H_0 : \beta_1 = 0\right)$$

Under $H_0$,
$$\frac{\hat{\beta}_1^2 \hat{\sigma}_X^2}{\hat{\sigma}_\varepsilon^2} \sim \frac{(n-2)}{(n-1)} * \frac{W_1 W_4}{W_2 W_3},$$

where $W_1 \sim \chi_1^2$, $W_2 \sim \chi_{n-1}^2$, $W_3 \sim \chi_{n-2}^2$ and $W_4 \sim \chi_{n-1}^2$, with $W_i$ s being independent. Explicit determination of critical values is still hard. The exact distribution of $T^2$ helps the task via simulations. The steps in the simulation work are detailed as follows.

Step 1: Fix $n$ and $\alpha$.
Step 2: Simulate $W_1 \sim \chi_1^2$.
Step 3: Simulate $W_2 \sim \chi_{n-1}^2$.
Step 4: Simulate $W_3 \sim \chi_{n-2}^2$.
Step 5: Simulate $W_4 \sim \chi_{n-1}^2$.
Step 6: Form the ratio $\frac{(n-2)}{(n-1)} * \frac{W_1 W_4}{W_2 W_3}$
Step 7: Repeat steps 2, 3, 4, 5 and 6 10,000 times.
Step 8: Calculate the $(1 - \alpha) * 100^{th}$ percentile $C_{n,\alpha}^2$ of the ratios.
Step 9: Calculate the positive square root $C_{n,\alpha}$ of $C_{n,\alpha}^2$.
Step 10: Repeat Step 9, 1000 times.
Step 11: Calculate mean and standard deviation of $C_{n,\alpha}$.
Step 12: Record the results average $(C_{n,\alpha}) \pm$ SD.

We are not solving for the critical value using an equation. The critical value is obtained via large scale simulations using the exact distribution of $T^2$ under the null hypothesis. It is normal to expect variation in the critical value from one simulation run to another. The standard deviation of the critical values emanating from Step 11 will capture the extent of variation. We have used R to run Steps 1 through 12. The code is provided in the supplement. For example, when n = 30, α = 0.05, average ($C_{n,\alpha}$) = 0.32634 with SD = 0.00320. When n = 45, α = 0.01, average ($C_{n,\alpha}$) = 0.37741 with SD = 0.00473.

One can also obtain the critical value $C_{n,\alpha}$ via the asymptotic distribution of T. If n is large, T ~ Normal (0, $\frac{(n-2)}{(n-3)(n-4)}$) approximately. There are several ways to establish asymptotic normality of T. The exact unconditional distribution of $\frac{\sigma_X}{\sigma}(\hat{\beta}_1 - \beta_1)\sqrt{n-1}$ is $t_{n-1}$, which is asymptotically N(0, 1). Then use the fact that $\hat{\sigma}_X$ is consistent for $\sigma_X$ and that $\hat{\sigma}$ consistent for σ. Since we know the variance of T exactly, we use this variance in the description of the asymptotic distribution of T. The critical values following the asymptotic distribution are identified.

| Level | Formula |
|---|---|
| 10% | $1.645 * \sqrt{\frac{n-2}{(n-3)(n-4)}}$ |
| 5% | $1.96 * \sqrt{\frac{n-2}{(n-3)(n-4)}}$ |
| 1% | $2.576 * \sqrt{\frac{n-2}{(n-3)(n-4)}}$ |

We have used R to run Steps 1 through 12. The code is provided in the supplement. In Table 1, we document the average critical value $C_{n,\alpha}$ for n = 20 (1) 100, α = 0.01, 0.05, and 0.10, along with the critical values coming from the normal approximation. The SDs are not included in the table. They are included in tables for sample sizes.
Table 1: Exact Critical Values for Levels 1%, 5%, and 10% and Normal Approximations

| samplesize | normal10 | criticalvalue10 | normal5 | criticalvalue5 | normal1 | criticalvalue1 |
|---|---|---|---|---|---|---|
| 20 | 0.423 | 0.417 | 0.504 | 0.518 | 0.663 | 0.75 |
| 21 | 0.41 | 0.404 | 0.488 | 0.501 | 0.642 | 0.721 |
| 22 | 0.398 | 0.392 | 0.474 | 0.486 | 0.623 | 0.697 |
| 23 | 0.387 | 0.382 | 0.461 | 0.472 | 0.606 | 0.674 |
| 24 | 0.376 | 0.372 | 0.449 | 0.459 | 0.59 | 0.655 |
| 25 | 0.367 | 0.363 | 0.437 | 0.447 | 0.575 | 0.634 |
| 26 | 0.358 | 0.355 | 0.427 | 0.436 | 0.561 | 0.618 |
| 27 | 0.35 | 0.346 | 0.417 | 0.426 | 0.548 | 0.602 |
| 28 | 0.342 | 0.339 | 0.408 | 0.417 | 0.536 | 0.587 |
| 29 | 0.335 | 0.332 | 0.399 | 0.408 | 0.525 | 0.572 |
| 30 | 0.329 | 0.326 | 0.391 | 0.399 | 0.514 | 0.56 |
| 31 | 0.322 | 0.32 | 0.384 | 0.391 | 0.505 | 0.548 |
| 32 | 0.316 | 0.314 | 0.377 | 0.384 | 0.495 | 0.536 |
| 33 | 0.311 | 0.308 | 0.37 | 0.376 | 0.486 | 0.525 |
| 34 | 0.305 | 0.303 | 0.364 | 0.37 | 0.478 | 0.515 |
| 35 | 0.3 | 0.298 | 0.357 | 0.363 | 0.47 | 0.504 |
| 36 | 0.295 | 0.293 | 0.352 | 0.357 | 0.462 | 0.496 |
| 37 | 0.291 | 0.289 | 0.346 | 0.352 | 0.455 | 0.487 |
| 38 | 0.286 | 0.284 | 0.341 | 0.346 | 0.448 | 0.479 |
| 39 | 0.282 | 0.28 | 0.336 | 0.341 | 0.441 | 0.471 |
| 40 | 0.278 | 0.276 | 0.331 | 0.336 | 0.435 | 0.464 |
| 41 | 0.274 | 0.272 | 0.326 | 0.33 | 0.429 | 0.456 |
| 42 | 0.27 | 0.269 | 0.322 | 0.326 | 0.423 | 0.45 |
| 43 | 0.267 | 0.265 | 0.318 | 0.322 | 0.418 | 0.443 |
| 44 | 0.263 | 0.262 | 0.314 | 0.318 | 0.412 | 0.437 |
| 45 | 0.26 | 0.259 | 0.31 | 0.314 | 0.407 | 0.431 |
| 46 | 0.257 | 0.255 | 0.306 | 0.31 | 0.402 | 0.425 |
| 47 | 0.254 | 0.252 | 0.302 | 0.306 | 0.397 | 0.42 |
| 48 | 0.251 | 0.249 | 0.299 | 0.303 | 0.393 | 0.414 |
| 49 | 0.248 | 0.246 | 0.295 | 0.298 | 0.388 | 0.409 |
| 50 | 0.245 | 0.244 | 0.292 | 0.296 | 0.384 | 0.404 |
| 51 | 0.242 | 0.241 | 0.289 | 0.292 | 0.38 | 0.399 |
| 52 | 0.24 | 0.239 | 0.286 | 0.289 | 0.376 | 0.395 |
| 53 | 0.237 | 0.236 | 0.283 | 0.286 | 0.372 | 0.39 |
| 54 | 0.235 | 0.234 | 0.28 | 0.283 | 0.368 | 0.386 |
| 55 | 0.233 | 0.231 | 0.277 | 0.28 | 0.364 | 0.382 |
| 56 | 0.23 | 0.229 | 0.274 | 0.277 | 0.361 | 0.378 |
| 57 | 0.228 | 0.227 | 0.272 | 0.275 | 0.357 | 0.374 |
| 58 | 0.226 | 0.225 | 0.269 | 0.272 | 0.354 | 0.369 |
| 59 | 0.224 | 0.223 | 0.267 | 0.269 | 0.35 | 0.366 |
| 60 | 0.222 | 0.221 | 0.264 | 0.267 | 0.347 | 0.362 |
| 61 | 0.22 | 0.219 | 0.262 | 0.264 | 0.344 | 0.359 |
| 62 | 0.218 | 0.217 | 0.26 | 0.262 | 0.341 | 0.356 |
| 63 | 0.216 | 0.215 | 0.257 | 0.26 | 0.338 | 0.353 |
| 64 | 0.214 | 0.213 | 0.255 | 0.26 | 0.335 | 0.349 |
| 65 | 0.212 | 0.211 | 0.253 | 0.255 | 0.332 | 0.346 |
| 66 | 0.211 | 0.21 | 0.251 | 0.253 | 0.33 | 0.343 |
| 67 | 0.209 | 0.208 | 0.249 | 0.251 | 0.327 | 0.34 |
| 68 | 0.207 | 0.206 | 0.247 | 0.249 | 0.324 | 0.337 |
| 69 | 0.206 | 0.205 | 0.245 | 0.247 | 0.322 | 0.334 |
| 70 | 0.204 | 0.203 | 0.243 | 0.245 | 0.319 | 0.331 |
| 71 | 0.202 | 0.202 | 0.241 | 0.243 | 0.317 | 0.329 |
| 72 | 0.201 | 0.2 | 0.239 | 0.241 | 0.315 | 0.326 |
| 73 | 0.199 | 0.199 | 0.238 | 0.24 | 0.312 | 0.324 |
| 74 | 0.198 | 0.197 | 0.236 | 0.238 | 0.31 | 0.321 |
| 75 | 0.197 | 0.196 | 0.234 | 0.236 | 0.308 | 0.319 |
| 76 | 0.195 | 0.194 | 0.233 | 0.234 | 0.306 | 0.316 |
| 77 | 0.194 | 0.193 | 0.231 | 0.233 | 0.304 | 0.314 |
| 78 | 0.192 | 0.192 | 0.229 | 0.231 | 0.301 | 0.312 |
| 79 | 0.191 | 0.191 | 0.228 | 0.23 | 0.299 | 0.309 |
| 80 | 0.19 | 0.19 | 0.226 | 0.228 | 0.297 | 0.307 |
| 81 | 0.189 | 0.188 | 0.225 | 0.226 | 0.295 | 0.305 |
| 82 | 0.187 | 0.187 | 0.223 | 0.225 | 0.294 | 0.303 |
| 83 | 0.186 | 0.186 | 0.222 | 0.224 | 0.292 | 0.301 |
| 84 | 0.185 | 0.184 | 0.22 | 0.222 | 0.29 | 0.299 |
| 85 | 0.184 | 0.183 | 0.219 | 0.221 | 0.288 | 0.297 |
| 86 | 0.183 | 0.182 | 0.218 | 0.219 | 0.286 | 0.295 |
| 87 | 0.182 | 0.181 | 0.216 | 0.218 | 0.284 | 0.293 |
| 88 | 0.181 | 0.18 | 0.215 | 0.217 | 0.283 | 0.291 |
| 89 | 0.179 | 0.179 | 0.214 | 0.215 | 0.281 | 0.289 |
| 90 | 0.178 | 0.178 | 0.213 | 0.214 | 0.279 | 0.287 |
| 91 | 0.177 | 0.177 | 0.211 | 0.213 | 0.278 | 0.286 |
| 92 | 0.176 | 0.176 | 0.21 | 0.211 | 0.276 | 0.284 |
| 93 | 0.175 | 0.175 | 0.209 | 0.21 | 0.275 | 0.282 |
| 94 | 0.174 | 0.174 | 0.208 | 0.209 | 0.273 | 0.28 |
| 95 | 0.173 | 0.173 | 0.207 | 0.208 | 0.272 | 0.279 |
| 96 | 0.172 | 0.172 | 0.205 | 0.207 | 0.27 | 0.278 |
| 97 | 0.171 | 0.171 | 0.204 | 0.205 | 0.269 | 0.276 |
| 98 | 0.171 | 0.17 | 0.203 | 0.204 | 0.267 | 0.274 |
| 99 | 0.17 | 0.169 | 0.202 | 0.203 | 0.266 | 0.272 |
| 100 | 0.169 | 0.168 | 0.201 | 0.202 | 0.264 | 0.271 |

Comments on Table 1:

   a. normal10 = critical value coming from the asymptotic distribution when α = 0.10.
   b. normal5 = critical value coming from the asymptotic distribution when α = 0.05.
   c. normal1 = critical value coming from the asymptotic distribution when α = 0.01.
   d. criticalvalue10 = critical value coming from the exact distribution of T when α = 0.10.
   e. criticalvalue5 = critical value coming from the exact distribution of T when α = 0.05.
   f. criticalvalue1 = critical value coming from the exact distribution of T when α = 0.01.
   g. When α = 0.10, |normal10 – criticalvalue10| ≤ 0.001 for n ≥ 50. One can enjoy the benefit of normal approximation when n ≥ 50.
   h. When α = 0.05, |normal5 – criticalvalue5| ≤ 0.001 for n ≥ 89. One can enjoy the benefit of normal approximation when n ≥ 89.
   i. For α = 0.01, Table 1 is not informative when |normal1 – criticalvalue1| ≤ 0.001.

## 5. Sample Size and Power

For a given level α, sample size n, and alternative value of $\beta_1 = A$, power is given by

$$\text{Power} = \Pr\left(\left|\frac{\hat{\beta}_1 * \hat{\sigma}_X}{\hat{\sigma}}\right| > C_{n,\alpha} \mid \beta_1 = A\right).$$

Suppose $1 - \beta$ is the specified power. For the required sample size, we set

$$1 - \beta = \Pr\left(\left|\frac{\hat{\beta}_1 * \hat{\sigma}_X}{\hat{\sigma}}\right| > C_{n,\alpha} \mid \beta_1 = A\right).$$

and solve for n. We will need the distribution of $\frac{\hat{\beta}_1 * \hat{\sigma}_X}{\hat{\sigma}}$, when $\beta_1 = A$. Rewrite

$$\frac{\hat{\beta}_1 * \hat{\sigma}_X}{\hat{\sigma}} = \frac{(\hat{\beta}_1 - \beta_1) * \hat{\sigma}_X}{\hat{\sigma}} + \frac{\beta_1 * \hat{\sigma}_X}{\hat{\sigma}}.$$

The distribution of $\frac{(\hat{\beta}_1 - \beta_1) * \hat{\sigma}_X}{\hat{\sigma}}$ (actually, the square of it) is described in Section 3 and it is free of the parameters of the regression model. Consequently, the random variables $\frac{(\hat{\beta}_1 - \beta_1) * \hat{\sigma}_X}{\hat{\sigma}}$ and $\frac{\beta_1 * \hat{\sigma}_X}{\hat{\sigma}}$ are independently distributed. Since $\hat{\sigma}$ and $\hat{\sigma}_X$ are independently distributed, $\left(\frac{\beta_1 * \hat{\sigma}_X}{\hat{\sigma}}\right)^2 \to^d \beta_1^2 * \frac{\sigma_X^2}{n-1} * W_5 * \frac{n-2}{\sigma^2} * \left(\frac{1}{W_6}\right) = \left(\beta_1 \frac{\sigma_X}{\sigma_\varepsilon}\right)^2 \left(\frac{n-2}{n-1}\right) W_5 W_6$, with $W_5 \sim \chi_{n-1}^2$, $W_6 \sim \chi_{n-2}^2$, and $W_5$ and $W_6$ being independent. An important fact emerges from these deliberations in that the distribution of $\frac{(\hat{\beta}_1 - \beta_1) * \hat{\sigma}_X}{\hat{\sigma}_\varepsilon} + \frac{\beta_1 * \hat{\sigma}_X}{\hat{\sigma}_\varepsilon}$ depends only on $\lambda = \beta_1 \frac{\sigma_X}{\sigma_\varepsilon}$, which we can deem as the effect size. In spite of all these labors, the distribution of $\frac{(\hat{\beta}_1 - \beta_1) * \hat{\sigma}_X}{\hat{\sigma}} + \frac{\beta_1 * \hat{\sigma}_X}{\hat{\sigma}_\varepsilon}$ is not amenable to direct and simple computations. For power calculations, we have resorted to simulations. We generate data from the regression model. Our strategy is as follows.

   a. Spell out $\lambda = \beta_1 \frac{\sigma_X}{\sigma}$.
   b. Take $\sigma_X = 1$ and $\sigma = 1$.
   c. For given n, draw a random sample $X_1, X_2, \ldots, X_n$ of size n from $N(0, 1)$.
   d. For each $1 \leq i \leq n$, draw a random sample of size 1 from $N(\beta_1 * X_i, 1)$.

e. We thus have the data: $(X_1, Y_1), (X_2, Y_2), \ldots, (X_n, Y_n)$ from the model $X \sim N(0, 1)$ and $Y | X \sim N(\beta_1 * X, 1)$. We are taking $\mu_X = 0$ and $\beta_0 = 0$. They do not play any role at all in the distributions identified. They can take any values.
f. Estimate $\beta_1$, $\sigma_X$, and $\sigma$.
g. Calculate $T = \frac{\hat{\beta}_1 * \hat{\sigma}_X}{\hat{\sigma}}$.
h. Check $|T| > C_{n,\alpha}$ with $C_{n,\alpha}$ coming from Table 1.
i. Set up a counter = 1 if $|T| > C_{n,\alpha}$, = 0, otherwise.
j. Repeat Steps a to i one-thousand times.
k. Calculate Power = $\frac{\# Counter=1}{1000}$.
l. If Power matches the targeted power $1 - \beta$, stop. Otherwise, keep experimenting with n until the targeted power is attained.
m. Once the sample size n is found out, we wanted to make sure that this is the right number. For the identified n, repeat Step a to l one thousand times. The average power and standard deviation is also reported.
n. Tables 2, 3, and 4 embody our effort under this strategy.
o. The R code is given in the supplement.

|  | $\beta^*(\sigma_x/\sigma_\varepsilon)$ | power | n | mean | sd |
|---|---|---|---|---|---|
| α=0.1 | 0.1 | 80% | 620 | 0.7993 | 0.013 |
|  |  | 90% | 870 | 0.9027 | 0.0095 |
|  |  | 95% | 1120 | 0.9546 | 0.0067 |
|  |  | 99% | 1690 | 0.993 | 0.0027 |
|  | 0.2 | 80% | 161 | 0.8259 | 0.0123 |
|  |  | 90% | 219 | 0.90007 | 0.0096 |
|  |  | 95% | 274 | 0.949 | 0.0069 |
|  |  | 99% | 440 | 0.9939 | 0.0024 |
|  | 0.3 | 80% | 73 | 0.8017 | 0.0124 |
|  |  | 90% | 100 | 0.9006 | 0.00995 |
|  |  | 95% | 124 | 0.9475 | 0.0071 |
|  |  | 99% | 195 | 0.931 | 0.0026 |
|  | 0.4 | 80% | 43 | 0.8031 | 0.0126 |
|  |  | 90% | 60 | 0.9073 | 0.0093 |
|  |  | 95% | 72 | 0.9518 | 0.0073 |
|  |  | 99% | 105 | 0.9896 | 0.0031 |
|  | 0.5 | 80% | 29 | 0.8045 | 0.0134 |
|  |  | 90% | 39 | 0.9003 | 0.0099 |
|  |  | 95% | 48 | 0.947 | 0.0068 |
|  |  | 99% | 69 | 0.989 | 0.0034 |
|  | 0.6 | 80% | 21 | 0.8 | 0.0129 |
|  |  | 90% | 28 | 0.8961 | 0.0096 |
|  |  | 95% | 35 | 0.9476 | 0.0069 |
|  |  | 99% | 52 | 0.9911 | 0.003 |

Table 2: Sample Size for Given Effect Size (ES), Power, Level of Significance 10%, mean of power in the validation step, and its standard deviation.

|  | $\beta^*(\sigma_x/\sigma_\varepsilon)$ | power | n | mean | sd |
|---|---|---|---|---|---|
| α=0.05 | 0.1 | 80% | 790 | 0.8006 | 0.0129 |
|  |  | 90% | 1080 | 0.9054 | 0.0088 |
|  |  | 95% | 1350 | 0.9557 | 0.0067 |
|  |  | 99% | 1850 | 0.9898 | 0.0032 |
|  | 0.2 | 80% | 199 | 0.797 | 0.0133 |
|  |  | 90% | 272 | 0.9039 | 0.0094 |
|  |  | 95% | 330 | 0.9497 | 0.0069 |
|  |  | 99% | 450 | 0.9891 | 0.0033 |
|  | 0.3 | 80% | 91 | 0.7978 | 0.0124 |
|  |  | 90% | 123 | 0.9028 | 0.0094 |
|  |  | 95% | 150 | 0.9505 | 0.0067 |
|  |  | 99% | 220 | 0.992 | 0.0028 |
|  | 0.4 | 80% | 53 | 0.7973 | 0.0128 |
|  |  | 90% | 70 | 0.8966 | 0.0096 |
|  |  | 95% | 87 | 0.9494 | 0.0071 |
|  |  | 99% | 121 | 0.9891 | 0.0034 |
|  | 0.5 | 80% | 36 | 0.8051 | 0.0124 |
|  |  | 90% | 48 | 0.9095 | 0.0091 |
|  |  | 95% | 58 | 0.95 | 0.0068 |
|  |  | 99% | 79 | 0.9888 | 0.0033 |
|  | 0.6 | 80% | 26 | 0.8005 | 0.0124 |
|  |  | 90% | 34 | 0.8985 | 0.0094 |
|  |  | 95% | 43 | 0.9547 | 0.0066 |
|  |  | 99% | 59 | 0.9901 | 0.0031 |

Table 3: Sample Size for Given Effect Size, Power, Level of Significance 5%, mean of power in the validation step, and its standard deviation.

| | $\beta^*(\sigma_x/\sigma_\varepsilon)$ | power | n | mean | sd |
|---|---|---|---|---|---|
| α=0.01 | 0.1 | 80% | 1180 | 0.8026 | 0.0124 |
| | | 90% | 1500 | 0.9015 | 0.0095 |
| | | 95% | 1760 | 0.946 | 0.0072 |
| | | 99% | 2440 | 0.9906 | 0.0031 |
| | 0.2 | 80% | 301 | 0.8045 | 0.0121 |
| | | 90% | 388 | 0.9046 | 0.0093 |
| | | 95% | 458 | 0.9529 | 0.0065 |
| | | 99% | 620 | 0.991 | 0.0031 |
| | 0.3 | 80% | 136 | 0.8044 | 0.0129 |
| | | 90% | 172 | 0.9012 | 0.0089 |
| | | 95% | 199 | 0.9432 | 0.0071 |
| | | 99% | 265 | 0.9872 | 0.0034 |
| | 0.4 | 80% | 78 | 0.8017 | 0.0124 |
| | | 90% | 95 | 0.8856 | 0.0099 |
| | | 95% | 118 | 0.949 | 0.007 |
| | | 99% | 158 | 0.9892 | 0.0033 |
| | 0.5 | 80% | 51 | 0.8042 | 0.0126 |
| | | 90% | 64 | 0.9011 | 0.0099 |
| | | 95% | 77 | 0.9464 | 0.007 |
| | | 99% | 104 | 0.9891 | 0.0032 |
| | 0.6 | 80% | 37 | 0.7975 | 0.0125 |
| | | 90% | 48 | 0.906 | 0.0089 |
| | | 95% | 56 | 0.9485 | 0.007 |
| | | 99% | 73 | 0.9874 | 0.0035 |

Table 4: Sample Size for Given Effect Size, Power, Level of Significance 1%, mean of power in the validation step, and its standard deviation.

Comments on Tables 2, 3, and 4:
   a. The first column in each table entertains three types of effect sizes: small (0.1, 0.2); medium (0.3, 0.4); and large (0.5, 0.6).
   b. The second column in each table lays the powers entertained.
   c. The third column in each table spells out the requisite sample size.
   d. The fourth column is the fruit of our effort to validate the sample size. At the ascertained sample size, data are generated under the specifications, power calculated, and power averaged over thousand times.
   e. The fifth column records the standard deviation of the thousand powers calculated.
   f. We are satisfied that the sample sizes laid out are holding true.

## 6. The Route via Sample Correlation Coefficient

The test statistic $T_1$ built upon the sample correlation coefficient $\hat{\rho}$ is another way to test the null hypothesis $\beta_1 = 0$, which is equivalent to $\rho = 0$. The statistic $T_1 \sim t_{n-2}$ under the null hypothesis. Under the alternative value $\rho$, $T_1 \sim$ non-central t with n-2 degrees of freedom with non-centrality parameter $\rho$. Sample size calculations are simple and direct. The software R has a package 'pwr,' which facilitates sample size calculations with specifications of the level of significance, power, and alternative value $\rho$. A table of sample sizes is given in Cohen (1988). Here, we want to explore connection between the sample sizes emanating from the test statistics T and T1. There is a connection between the correlation coefficient and effect size λ.

$$\rho = \frac{\beta_1 * \sigma_X}{\sqrt{\beta_1^2 \sigma_X^2 + \sigma^2}} = \frac{1}{\sqrt{1 + \frac{1}{\lambda^2}}} \quad (4)$$

A graph of the relationship is presented in Graph1.

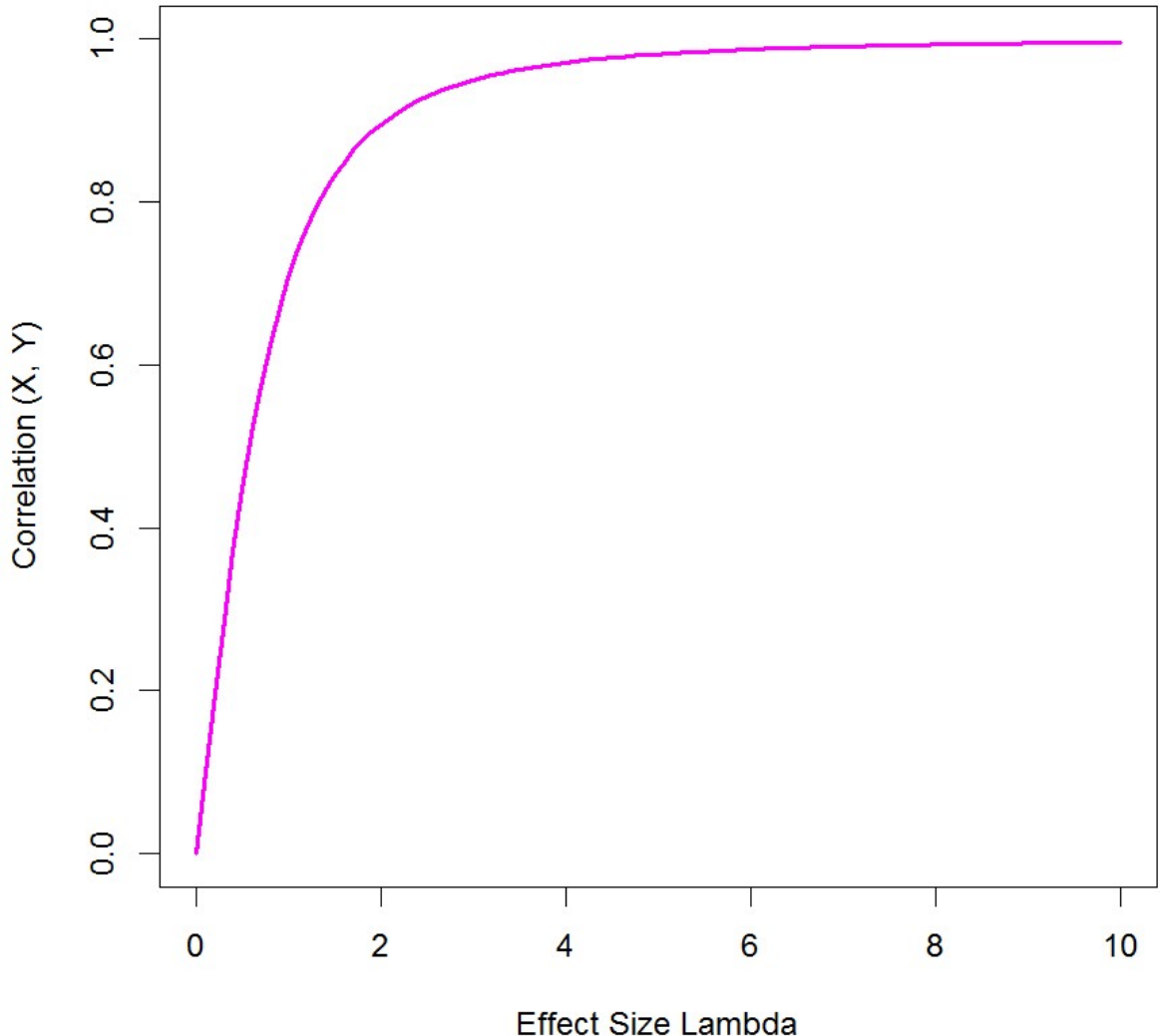

Graph1: Graph of ρ = 1/sqrt(1 + 1/λ²)

We now have two sources for sample sizes. In the environment of the slope parameter, we will spell out the level of significance, power, and effect size λ. We will record the sample size. For the given effect size λ, we will calculate the matching correlation coefficient (4), and then determine the required sample size. In the next section, we will contrast the sample sizes.

7. **Contrasting sample sizes**

**In the first table, when the level is 10%, we contrast sample sizes coming from two sources, one based on the test using the slope parameter and effect size λ, and the other based on the matching correlation coefficient.**

**In the second table, the level entertained is 5%, and in the third, it is 1%.**

| Level | ES=β*(σₓ/σ) | corr=ρ | power | Sample Size | | difference |
|---|---|---|---|---|---|---|
| | | | | Slope Test | CorrTest | |
| α=0.1 | 0.1 | 0.0995 | 80% | 620 | 622 | -2 |
| | | | 90% | 870 | 861 | 9 |
| | | | 95% | 1120 | 1088 | 32 |
| | | | 99% | 1690 | 1584 | 106 |
| | 0.2 | 0.1961 | 80% | 161 | 159 | 2 |
| | | | 90% | 219 | 219 | 0 |
| | | | 95% | 274 | 276 | -2 |
| | | | 99% | 440 | 401 | 9 |
| | 0.3 | 0.2873 | 80% | 73 | 73 | 0 |
| | | | 90% | 100 | 100 | 0 |
| | | | 95% | 124 | 126 | -2 |
| | | | 99% | 195 | 182 | 7 |
| | 0.4 | 0.3714 | 80% | 43 | 43 | 0 |
| | | | 90% | 60 | 58 | 2 |
| | | | 95% | 72 | 73 | -1 |
| | | | 99% | 105 | 106 | -1 |
| | 0.5 | 0.4472 | 80% | 29 | 29 | 0 |
| | | | 90% | 39 | 39 | 0 |
| | | | 95% | 48 | 49 | -1 |
| | | | 99% | 69 | 70 | -1 |
| | 0.6 | 0.5145 | 80% | 21 | 21 | 0 |
| | | | 90% | 28 | 29 | -1 |
| | | | 95% | 35 | 36 | -1 |
| | | | 99% | 52 | 51 | 1 |

**Table 5: Contrasting Sample Sizes Slope vs Correlation for 10% Level**

| Level | ES=β*(σₓ/σ) | corr=ρ | power | Sample Size | | difference |
|---|---|---|---|---|---|---|
| | | | | Slope Test | CorrTest | |
| α=0.05 | 0.1 | 0.0995 | 80% | 790 | 790 | 0 |
| | | | 90% | 1080 | 1057 | 23 |
| | | | 95% | 1350 | 1306 | 44 |
| | | | 99% | 1850 | 1846 | 4 |
| | 0.2 | 0.1961 | 80% | 199 | 201 | -2 |
| | | | 90% | 272 | 269 | 3 |
| | | | 95% | 330 | 332 | -2 |
| | | | 99% | 450 | 468 | -18 |
| | 0.3 | 0.2873 | 80% | 91 | 92 | -1 |
| | | | 90% | 123 | 123 | 0 |
| | | | 95% | 150 | 151 | -1 |
| | | | 99% | 220 | 213 | 7 |
| | 0.4 | 0.3714 | 80% | 53 | 54 | -1 |
| | | | 90% | 70 | 72 | -2 |
| | | | 95% | 87 | 88 | -1 |
| | | | 99% | 121 | 123 | -2 |
| | 0.5 | 0.4472 | 80% | 36 | 37 | -1 |
| | | | 90% | 48 | 48 | 0 |
| | | | 95% | 58 | 59 | -1 |
| | | | 99% | 79 | 82 | -3 |
| | 0.6 | 0.5145 | 80% | 26 | 27 | -1 |
| | | | 90% | 34 | 35 | -1 |
| | | | 95% | 43 | 43 | 0 |
| | | | 99% | 59 | 59 | 0 |

**Table 6: Contrasting Sample Sizes Slope vs Correlation for 5% Level**

| Level | ES=β*(σₓ/σ) | corr=ρ | power | Sample Size | | difference |
|---|---|---|---|---|---|---|
| | | | | Slope Test | CorrTest | |
| α=0.01 | 0.1 | 0.0995 | 80% | 1180 | 1175 | 5 |
| | | | 90% | 1500 | 1496 | 4 |
| | | | 95% | 1760 | 1790 | -30 |
| | | | 99% | 2440 | 2414 | 26 |
| | 0.2 | 0.1961 | 80% | 301 | 299 | 2 |
| | | | 90% | 388 | 380 | 8 |
| | | | 95% | 458 | 454 | 4 |
| | | | 99% | 620 | 612 | 8 |
| | 0.3 | 0.2873 | 80% | 136 | 137 | -1 |
| | | | 90% | 172 | 173 | -1 |
| | | | 95% | 199 | 207 | -8 |
| | | | 99% | 265 | 278 | -13 |
| | 0.4 | 0.3714 | 80% | 78 | 80 | -2 |
| | | | 90% | 95 | 101 | -6 |
| | | | 95% | 118 | 120 | -2 |
| | | | 99% | 158 | 161 | -3 |
| | 0.5 | 0.4472 | 80% | 51 | 53 | -2 |
| | | | 90% | 64 | 67 | -3 |
| | | | 95% | 77 | 80 | -3 |
| | | | 99% | 104 | 107 | -3 |
| | 0.6 | 0.5145 | 80% | 37 | 39 | -2 |
| | | | 90% | 48 | 49 | -1 |
| | | | 95% | 56 | 58 | -2 |
| | | | 99% | 73 | 77 | -4 |

**Table 7: Contrasting Sample Sizes Slope vs Correlation for 1% Level**

Two key features emerge from the tables. 1. At the low effect size 0.1, sample sizes do differ substantially at 95% and 99% powers. 2. In the remaining cases, a good agreement between the sample sizes prevails. Within the purview of these scenarios, test hopping is feasible. If one is given the level α, power 1-β, effect size λ, calculate the matching correlation coefficient ρ. Determine the sample size required based on the built upon the sample correlation coefficient and proffer it as the required sample size for test based on the sample slope. Of course, make sure that we are with in the purview of the scenarios discussed.